\documentclass[%
 prd,reprint,
showpacs,preprintnumbers,
nofootinbib,
 amsmath,amssymb,
 aps,
]{revtex4-1}

\usepackage{graphicx}
\usepackage{dcolumn}
\usepackage{bm}
\usepackage{nicefrac}
\usepackage{breqn}

\usepackage[usenames,dvipsnames]{xcolor}

\newcommand{\Slash}[1]{\ooalign{\hfil/\hfil\crcr$#1$}}

\newcommand{\flaligne}[1]{\begin{flalign} #1 \end{flalign}}
\newcommand{\bracket}[1]{\left( #1 \right)}

\begin{document}


\title{Elucidating the effect of intermediate resonances in the quark interaction kernel 
  on the time-like electromagnetic pion form factor } 

\makeatletter
\let\cat@comma@active\@empty
\makeatother

\author{\'Angel S. Miramontes L\'opez}
\email{angel-aml@hotmail.com}
\affiliation{Instituto de F\'isica y Matem\'aticas, Universidad Michoacana de San Nicol\'as de Hidalgo, Morelia, Michoac\'an 58040, Mexico}

\author{H\`elios Sanchis Alepuz}
\email{helios.sanchis-alepuz@silicon-austria.com \\ hsanchisalepuz@gmail} 
\affiliation{Silicon Austria Labs GmbH, Inffeldgasse 25F, 8010 Graz, Austria}

\author{Reinhard Alkofer}
\email{reinhard.alkofer@uni-graz.at}
\affiliation{Institute of Physics, University of Graz, NAWI Graz, Universit\"atsplatz 5, 8010 Graz, Austria}

\date{\today}

\begin{abstract}
An exploratory study of the time-like pion electromagnetic form factor  in a Poincar\'e-covariant bound state 
formalism in the isospin symmetric limit is presented. Starting from a quark interaction kernel representing
gluon-intermediated interactions for valence-type quarks, non-valence effects are included by introducing 
pions as explicit degrees of freedom. The two most important qualitative aspects are, in view of the presented 
study, the opening of the dominant $\rho$-meson decay channel and the presence of a multi-particle 
branch cut setting in when the two-pion threshold is crossed. Based on a recent respective computation 
of the quark-photon vertex, the pion electromagnetic form factor for space-like and time-like
kinematics is calculated. The obtained results for its absolute value and its phase compare favorably 
to the available experimental data, and they are analyzed in detail by confronting them to the expectations
based on an isospin-symmetric version of a vector-meson dominance model. 
\end{abstract}

\pacs{11.10.St, 13.30.Eg, 13.40.Gp, 14.40.n}
\maketitle

\section{\label{sec:intro}Introduction}

Hadronic time-like form factors will be measured in upcoming experiments to an unprecedented precision.
An understanding of these quantities which are displaying pronounced structures originating from hadron
resonances will contribute significantly to our knowledge on the relation between the hadrons' substructure 
and the hadron spectrum. On the one hand, based on respective experimental and theoretical progress in the
last decades, it is by now evident that hadron resonances can be, at least in principle, described in terms of quarks 
and gluons. The latter, being the QCD degrees of freedom, are considered to be complete in the sense that 
they allow for a computation of every hadronic observable. Changing the perspective in an attempt to understand 
the Strong Interaction starting from the low-energy regime, a possible way of phrasing the phenomenon of confinement 
in QCD is the statement that all possible hadronic degrees of freedom will form also a complete set of physical states. 
Therefore, the equivalence of descriptions of observables in either quark and glue or hadronic degrees of freedom 
is a direct consequence of confinement and unitarity. This is the gist of a chain of arguments which can be 
sophisticated and applied to many different phenomena involving hadrons. The related picture is known under the 
name ``quark-hadron duality'', and its consequences have been verified on the qualitative as well as the 
semi-quantitative level, for a review see, e.g., Ref.\ \cite{Melnitchouk:2005zr}. A verification of this duality is, beyond 
the trivial fact of the absence of coloured states, the clearest experimental signature for confinement. To appreciate the 
scope of such a scenario it is important to note that a perfect orthogonality of the quark-glue degrees of freedom 
on the one hand and hadronic states on the other hand, and thus the perfect absence of ‘‘double-counting’’ in any of 
the two ‘‘languages’’, is nothing else but another way to express confinement.

Gaining insight into the interplay between formation of hadronic bound states, consisting of quarks and gluons, and 
the open decay channels of the respective resonance is an essential element of every study of time-like form factors
in kinematic regions close to a resonance. Here, attention should be paid to the fact that the hadron whose form
factor is investigated and the hadronic resonance which is apparent in the form factor are both to be described
as composite objects of quarks and gluons. The  same is true for the hadronic decay products in a hadronic or 
semi-leptonic decay of the resonance. This makes evident that an approach to calculate a time-like form factor from
QCD, or from a microscopic model based on QCD degrees of freedom, faces the challenging task to treat all elements
appearing in the calculation on the same footing and to a sufficient degree of sophistication if the result is intended to allow for conclusions on the dynamics underlying such a form factor. 

Herein, we will report on an exploratory study of the 
time-like pion electromagnetic form factor using functional methods. More precisely, we will employ a combination 
of Bethe-Salpeter and Dyson-Schwinger equations (for recent reviews on this and related approaches see, {\it e.g.}, \cite{Cloet:2013jya, Eichmann:2016yit,Huber:2018ned,Sanchis-Alepuz:2017jjd}). Although such an approach is 
capable of allowing a first-principle calculation (see, {\it e.g.}, the computation of the glueball spectrum reported 
in ref.\ \cite{Huber:2020ngt}), for the task at hand this is yet out of reach. 
To grasp all essential features of the pions' time-like form factor\footnote{There are several investigations of the 
space-like pion electromagnetic form factor in the Dyson-Schwinger--Bethe-Salpeter approach, early examples 
include \cite{Langfeld:1989en,Roberts:1994hh,Maris:2000sk}. A calculation of this form factor spanning the entire
domain of space-like momentum transfers is described in ref.\ \cite{Chang:2013nia}. }
 one needs to describe at least 
(i) the pion as bound state of quark and antiquark thereby at the same time taking 
into account its special role as would-be Goldstone boson of the dynamically broken chiral symmetry of QCD, 
(ii) the mixing, respectively, the interference of the $\rho$-meson, being described also as a quark-antiquark bound 
state, with a virtual photon when this photon is in turn coupled to a quark-antiquark pair via the fully renormalized 
quark-photon vertex, and 
(iii) the dominant decay channel of the $\rho$-meson, namely, $\rho \to \pi \pi$. The study presented here is now in 
two aspects exploratory. First of all, the interaction between quarks and antiquarks is modelled in such a way that 
the essential features, as implied by QCD and phenomenology, are taken into account but that it is on the other hand
still manageable in such an involved calculation. Second, in several places we will make technical simplifications, 
especially when the such introduced error can for good reasons assumed to be small and the reduction in the 
computing time needed is substantial. We thus aim here more for an understanding of how the different features of the form 
factor arise from the QCD degrees of freedom than for a quantitative agreement with the experimental data.

In the chosen model for the quark-antiquark interaction, besides a gluon-mediated interaction also pions will be
included explicitly. The reason for this is as follows: if one were able to take into account the fully renormalized 
quark-gluon vertex exactly within this approach, hadronic degrees of freedom will effectively emerge and 
thus be included in the interaction between quarks and antiquarks, respectively, they 
will back-feed on the quarks' dynamics.   Due to the pions' Goldstone boson nature,
and especially due to the implied small pion mass, the pions are the most important low-energy degrees of 
freedom within the Strong Interaction, as, {\it e.g.}, also elucidated by chiral perturbation theory. And as 
in order to describe the physics of decays non-valence effects need to be taken into account, it is some minimal 
requirement for the investigation reported here to include pions as the most important non-valence-type interaction 
mediator in the sub-GeV region. 

The interaction model herein is also chosen in view of a possible generalisation to the study of baryon form factors,
and hereby especially the nucleons' time-like form factor. In this respect one can build on existing
calculations of space-like form factors from bound state amplitudes, see, {\it e.g.}, refs.\ 
\cite{Eichmann:2016yit,Nicmorus:2010sd,Eichmann:2011vu,Eichmann:2011pv,Sanchis-Alepuz:2013iia,Sanchis-Alepuz:2017mir}
for some recent respective work. A thorough understanding of the proton time-like form factor at very low $Q^2$ is 
a very timely subject as the upcoming PANDA experiment possesses the unique possibility to measure the proton's
electromagnetic form factors in the so-called unphysical region through the process $\bar p p \to l^+l^-\pi^0$, $l=e, \mu$
\cite{Fischer:2021kcr}. At large $Q^2$ the question of the onset of the convergence scale between the space-like and
the time-like form factors arises. 

However, also the pions' time-like electromagnetic form factor will be studied further by upcoming experiments, among other reasons because recently the consistency of the available data sets has been questioned \cite{Ananthanarayan:2020vum}. Based on the long-known fact that the $\tau$ radiative decay allows to extract the 
pion form factor \cite{Kim:1979hx} and that a very large number of $\tau$-leptons are produced at B-meson factories
further high-precision data in sub-GeV region will become available.\footnote{In this work we will compare to the dataset of ref.\ \cite{Akhmetshin:2006bx} available at https://www.hepdata.net/record/ins728302.}

Besides earlier lattice QCD calculations of the pions' space-like electromagnetic form factor 
\cite{Brommel:2006ww,Frezzotti:2008dr,Boyle:2008yd,Aoki:2009qn,Nguyen:2011ek,Brandt:2013dua,Koponen:2013boa}
recently calculations of the time-like pion form factor have become available
\cite{Meyer:2011um,Feng:2014gba,Bulava:2015qjz,Erben:2019nmx}. These results typically show a good 
agreement with the experimental data. In those calculations, the extraction of the time-like form factor employs a parameterisation 
based on the vector meson dominance (VMD) picture to extract from the lattice data at discrete energies the time-like 
pion form factor.

The exploratory calculation presented herein is done in the isospin symmetric limit. One of the effects of isospin 
breaking clearly visible in the time-like pion form factor is $\rho$-$\omega$ mixing, see, {\it e.g.}, the review 
\cite{OConnell:1995nse}. To this end we will employ the VMD-based fit given in \cite{OConnell:1995nse} and modify 
it such that an expected form of the pion form factor without this mixing effect is extracted.\footnote{In 
ref.\ \cite{Jegerlehner:2011ti} another method has been used to remove the $\rho$-$\omega$ mixing effects from the data.}
Furthermore, we will present a simplified but numerically quite accurate VMD parameterisation of the time-like form 
factor which then serves as a basis for a detailed analysis of our results. Here, the focus is more on a comparison of 
our results with the form expected on the basis of the VMD parameterisation than on numerical agreement.\footnote{A precise representation of the time-like form factor requires parameterisations including excited $\rho$-mesons, 
respective examples can be found in refs.\ \cite{Schael:2005am,Davier:2005xq}. Two-photon effects, on the other 
hand, can very likely be safely neglected, for a corresponding study of the form factor at large 
momentum transfer see \cite{Chen:2018tch}.} 

This paper is organized as follows: In Sect.\ \ref{sec:PionFF} we review some facts about the electromagnetic 
pion form factor, 
employ the VMD-based fit given in \cite{OConnell:1995nse}  to remove the $\rho$-$\omega$ mixing effects, and 
provide an expected form of the pion form factor.
In Sect.\ \ref{sec:DSE_BSE} we present our approach based on Bethe-Salpeter and Dyson-Schwinger equations.
Our results are presented and analyzed in Sect.~\ref{sec:results}. In Sect.~\ref{sec:outlook} we present conclusions and 
an outlook. Some technical details are deferred to two appendices.

\section {\label{sec:PionFF} The time-like electromagnetic pion form factor}

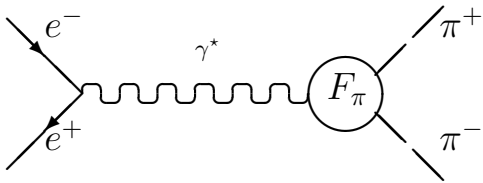
\begin{figure}[ht]
\begin{center}
\setlength{\unitlength}{0.5mm}
\begin{picture}(150,60)
\thicklines
\multiput(32.5,30)(10,0){6}{\oval(5,5)[t]}
\multiput(37.5,30)(10,0){6}{\oval(5,5)[b]}
\put(60,40){$\gamma^\star $}
\put(10,50){\vector(1,-1){10}}
\put(20,40){\line(1,-1){10}}
\put(20,45){\Large $e^-$}
\put(30,30){\vector(-1,-1){10}}
\put(20,20){\line(-1,-1){10}}
\put(20,15){\Large $e^+$}
\put(100,30){\circle{20}}
\put(95,28.5){\Large $F_\pi$}
\put(125,45){\Large $\pi^+$}
\put(125,15){\Large $\pi^-$}
\multiput(108,35)(10,10){2}{\line(1,1){8}}
\multiput(108,25)(10,-10){2}{\line(1,-1){8}}
\end{picture}
\end{center}
\caption{Electron-positron pair annihilating to a virtual photon with time-like momentum which then 
decays to a pion pair. } 
\label{Fig:PiFF}     
\end{figure}

The pion, being a composite object, does not have a point-like interaction with the electromagnetic field, and 
the related substructure, the pion being a pseudoscalar, is related to one form factor. Considering, for example, 
the scattering of an electron off a pion $\pi^+$ one can describe the leptonic part of the interaction quite
precisely in lowest-order perturbation theory, {\it i.e.}, one  considers the process in which the electron emits a 
virtual photon, and the latter couples to the pion. Defining the form factor $F_\pi (s)$ via the relation
\begin{equation}
\langle \pi^+(p_1) |  j^\mu _{e.m.} |  \pi^+(p_2) \rangle = e (p_1+p_2)^\mu F_\pi (q^2) \, , 
\end{equation}
where $q^\mu = p_1^\mu - p_2^\mu$ is the virtual photon momentum,
and $e$ is the elementary electric charge. The $S$-matrix element for electron-pion scattering is then 
proportional to the form factor,
\begin{eqnarray}
i {\cal M}_{e\pi \to e \pi } &=&  e (p_1+p_2)^\mu F_\pi (q^2) D_{\rm photon}^{\mu \nu } (q) 
\nonumber \\ && \qquad
\left( ie \bar u (k_1,s_1) \gamma _\nu u(k_2,s_2) \right) \, ,
\end{eqnarray}
where $D_{\rm photon}^{\mu \nu } (q)$ is the photon propagator, and $u(k_i,s_i)$ is the electron spinor,
see, {\it e.g.}, Sect.\ 8.4 of  \cite{Aitchison:2003tq} for more details. The kinematics of this 
scattering process is such that 
the photon momentum is space-like, and without loss of generality one can assume the form factor 
$F_\pi (s)$ to be real. Charge conservation requires that for a real photon one has
$ F_\pi (0) = 1$.

Turning to the process of electron-positron annihilation into a pion pair the corresponding $S$-matrix 
element is again proportional to the form factor, see , {\it e.g.}, Sect.\ 8.5 of  \cite{Aitchison:2003tq},
\begin{eqnarray}
i {\cal M}_{e^+e^-\to \pi^+ \pi^- } &=&  e (p_1+p_2)^\mu F_\pi (q^2) D_{\rm photon}^{\mu \nu } (q) 
\nonumber \\ && \qquad
\left(- ie \bar v (k_1,s_1) \gamma _\nu u(k_2,s_2) \right) \, ,
\label{eq:Smatrix}
\end{eqnarray}
with appropriately redefined momenta. Especially, the virtual photon momentum is now time-like, 
{\it cf.} Fig.\ \ref{Fig:PiFF}, and 
one measures in such an annihilation process to a pion pair the form factor for time-like momenta. Above
the two-pion production threshold, {\it i.e.}, in the physical region,  the corresponding cut in the amplitude  \eqref{eq:Smatrix} necessitates to treat the time-like pion form factor as a complex quantity, it fulfils the
dispersion relation 
\begin{equation}
F_\pi (q^2) = 1 + \frac {q^2}\pi \int _{4m_\pi^2} ds \frac {{\cal I}m \,   F_\pi (s) }{s(s-q^2-i \epsilon )} \, . 
\label{eq:DispRel}
\end{equation}
Especially, it is expected that the phase of the pion form factor varies strongly in a two-pion resonance 
region. Below the inelastic threshold, {\it i.e.}, for $s< 4 m_\pi^2$,
 the time-like pion form factor is, via Watson's final state theorem,
related to the isovector P-wave scattering phase shift $\delta_{1,1} (s) $:
\begin{eqnarray}
{\cal I}m \,    F_\pi (q^2) &=& \frac 1 {2i} \left( 1 - e^{-2i \delta_{1,1}(s)} \right)  F_\pi (s+i \epsilon ) 
\nonumber \\
& = & \sin (\delta_{1,1}(s)) e^{-i \delta_{1,1}(s) }  F_\pi (s+i \epsilon ) \, .
\end{eqnarray}

As depicted in Fig.\  \ref{Fig:PiFF}, the pion form factor contains, for time-like as well as for space-like 
photon virtualities, all kind of interaction processes turning a photon into a pion pair. As strong-interaction
processes dominate the corresponding amplitude, and as gluons do not couple directly to photons, one
decisive element of the pion form factor is the amplitude describing how the photon couples to a quark,
taking hereby all possible contributing QCD processes into account.  This amplitude is exactly the 
full quark-photon vertex. As we will see in the following, this correlation function carries information about 
the virtual photon's hadronic substructure. And as will be described in detail in the next section, 
the other quantities needed then for calculating the pion 
form factor are the fully renormalized quark propagator and the pion bound state amplitude, 
the latter describing how an antiquark and a quark  form a pionic bound state.

While our calculation is based on QCD degrees of freedom it is capable of providing an understanding 
why in the resonance region a vector-meson dominance (VMD) picture provides very good results for the 
time-like pion form factor. 
In addition, it will elucidate to which extent a VMD picture might be applicable in other kinematic regions.

In the VMD picture, the hadronic contribution to the photon propagator is given by mixing with 
electrically neutral vector mesons. Restricting to light-quark mesons, the corresponding vector meson 
is the $\rho^0$, {\it i.e.}, the uncharged member of the isotriplet of vector mesons. In a would-be isospin
symmetric world this would be the only vector meson below one GeV with which a virtual photon 
mixes because the isosinglet $\omega$ will not mix with the photon due to $G$-parity.

In the real world, isospin symmetry is broken, and one of the many effects of isospin breaking is 
$\rho$-$\omega$ mixing, see, {\it e.g.}, the review \cite{OConnell:1995nse}. 
This mixing and the resulting interference of states lead 
to quite some pronounced structure in the time-like pion form factor around 
$m_\omega^2$. Indeed, it is by now well understood that the sharp dip in the experimental data stems from a combination of $\rho$-$\omega$ mixing and interference effects between the decays 
$\rho\rightarrow\pi\pi$ and $\omega\rightarrow \pi\pi$, the latter being isospin breaking, see, {\it e.g.}, 
Ref.\ \cite{OConnell:1995nse}. Moreover, it has been estimated that this combination of mixing and interference effects decreases the height of the bump of the order of up to $10\%$ (see \cite{Jegerlehner:2011ti} and references therein).

\begin{figure}[ht]
\includegraphics[width=0.49\textwidth]{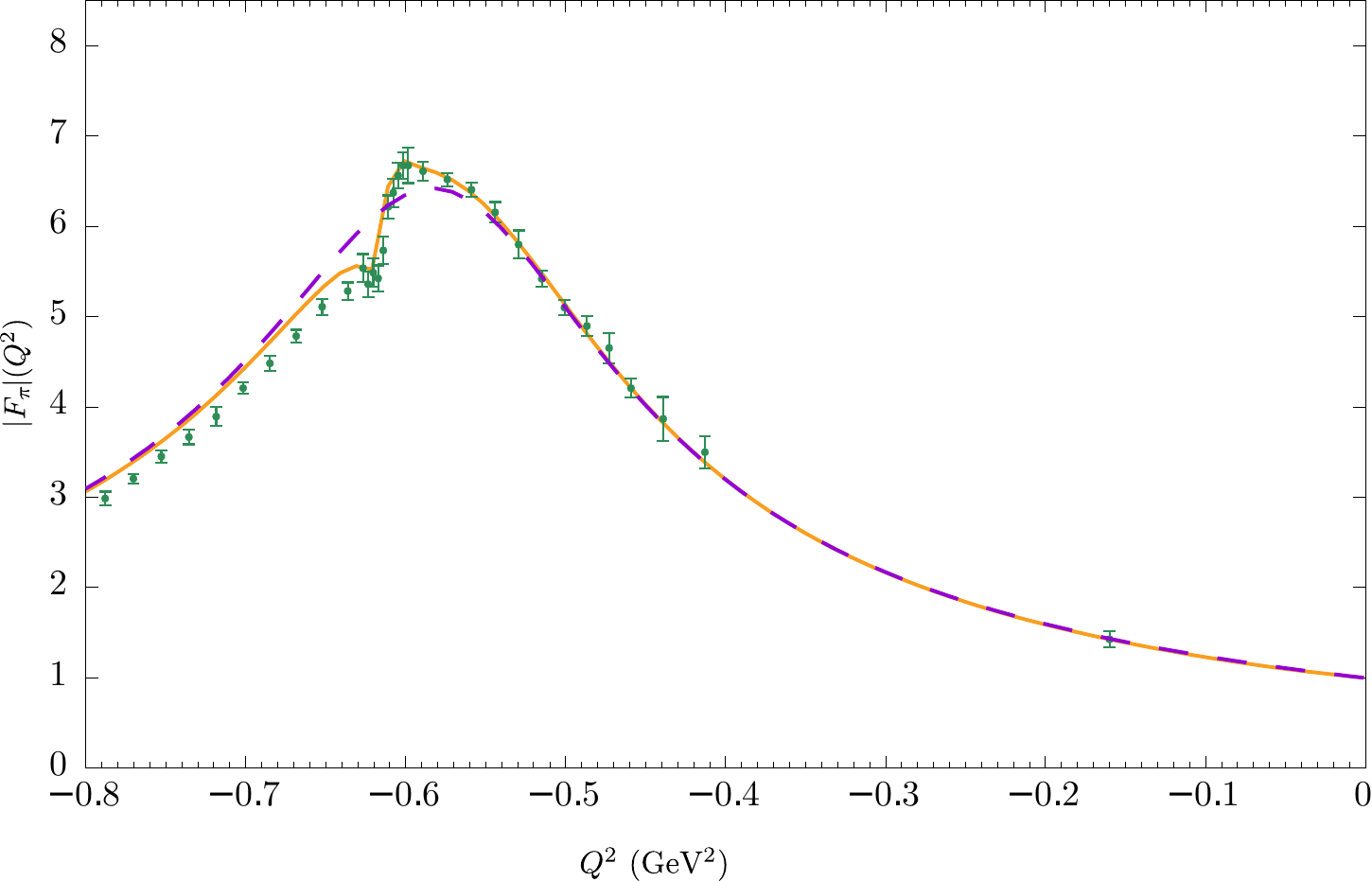}
\caption{Absolute value of the pion form factor in the time-like ($Q^2<0$) domain from the VMD-based
fit given in ref.\ \cite{OConnell:1995nse} (full line) in comparison to the experimental data \cite{Ananthanarayan:2020vum}. 
The dashed line is based on the same expression
but the mixing matrix element put to zero, $\Pi_{\rho\omega}=0$. }
\label{fig:FpiVMDplot}    
\end{figure}

As the here presented exploratory calculation is performed in the isospin
limit\footnote{Isospin breaking and thus the effect of $\rho$-$\omega$ mixing on the pion form
factor  is currently under investigation, the corresponding results will be published elsewhere.}
and thus $\rho$-$\omega$ mixing is neglected
we estimate its effect by comparing the VMD-based fit to the pion form
factor given in ref.\ \cite{OConnell:1995nse} with a plot of the same expression but the mixing 
matrix element put to zero, $\Pi_{\rho\omega}=0$, see Fig.\ \ref{fig:FpiVMDplot}. 
As expected the resulting curve is much smoother 
than the one including the $\rho$-$\omega$ interference effect. Around the $\omega$ mass the deviation 
of the two curves can be as large as almost ten percent, however, this effect is limited to a small 
interval, and the two curves are practically indistinguishable outside this small interval. Therefore 
we expect our calculation to reproduce all qualitative features of the curve representing the case 
with $\rho$-$\omega$ mixing switched off, and to be in a reasonable quantitative agreement with it.

It is instructive to analyze the momentum behaviour of a simplified version of the fit given in ref.\ \cite{OConnell:1995nse}. 
In order to be in agreement with the notation in the following presentation we
introduce the photon virtuality with the convention that $Q^2<0$ corresponds to the time-like region.  
In the above used fit a momentum dependent expression for the width of the $\rho$-meson is used.
It displays the two-pion cut,
 \begin{equation}
\Gamma_\rho (Q^2) \propto  \left( -Q^2 - 4 m_\pi^2 \right)^{3/2} \Theta ( -Q^2 - 4 m_\pi^2)\, , 
\end{equation}
which is important for a qualitatively correct analytic structure of the pion form factor. However, especially close 
to the maximum of the pion form factor, the pion mass is quantitatively negligible, and for vanishing 
pion mass the momentum dependent width assumes for time-like $Q^2<0$ the relatively simple form
\begin{equation}
\Gamma_\rho (Q^2) = \bar  \Gamma_\rho  |Q^2| / m_\rho^2\, , 
\end{equation}
where $  \bar \Gamma_\rho = \Gamma_\rho (-m_\rho^2)$. This then leads to the simplified but still quite accurate 
form for the fit\footnote{In contrast to a constant width approximation the pole is in this parameterization and thus in the 
employed fit not located at
$Q^2=-m_\rho^2+i m_\rho \bar \Gamma_\rho $ but at 
$Q^2=(-m_\rho^2+i m_\rho \bar \Gamma_\rho)/(1+\bar \Gamma_\rho^2/m_\rho^2) $, {\it i.e.}, real and imaginary
part of the pole position are decreased by 3.7\% \ if the same values for the $\rho$'s mass and width are used.}
\begin{eqnarray}
F_\pi (Q^2+i\epsilon ) &=& 1 - \frac{g_{\rho\pi\pi}}{g_\rho} \frac {Q^2(Q^2+m_\rho^2)}{(Q^2+m_\rho^2)^2 + Q^4 
\bar \Gamma^2 / m_\rho^2 } \nonumber \\
&+& i  \frac{g_{\rho\pi\pi}}{g_\rho} \frac {Q^4 \bar \Gamma / m_\rho }
{(Q^2+m_\rho^2)^2 + Q^4  \bar \Gamma^2 / m_\rho^2 } \, ,
\label{eq:VMDsimplified}
\end{eqnarray}
where $\epsilon \to 0^+$ has been introduced to fix the sign of the imaginary part. Hereby, the coupling constant
$g_{\rho\pi\pi}$ is determined from $\Gamma _{\rho\pi\pi} = \bar \Gamma = 149$ MeV to be 
$g_{\rho\pi\pi} \approx 6$, and $g_\rho$ is a parameter reflecting the strength of the 
$\gamma$-$\rho^0$-mixing, which is then described by the effective Lagrangian
$$
{\cal L}_{\rho \gamma} = - \frac {e\, m_\rho^2}{g_\rho} \rho_\mu^0 A^\mu \, .
$$
From the partial width $\Gamma _{\rho e^+e^-} = 7$ keV one infers $g_\rho \approx 5$.

The form given in Eq.\ \eqref{eq:VMDsimplified} motivates to compare the obtained results for the real and the 
imaginary part of the time-like form factor from our calculation to a rational, resp., Pad\'e fit. Phrased otherwise,
 Eq.\ \eqref{eq:VMDsimplified} formalises the expectation for the time-like form factor based on 
the VMD picture, and our results based on the quark-photon vertex function and the pion bound state 
amplitude will be analysed by discussing how much they extend beyond to this form.

\section{\label{sec:DSE_BSE}Dyson-Schwinger and Bethe-Salpeter  formalism}

We determine the necessary input for the calculation of the pion electromagnetic form factor using a combination of 
Bethe-Salpeter (BSE) and Dyson-Schwinger equations (DSE). To make this presentation 
self-contained we summarise in this section the most relevant aspects of the approach, for more details see, 
{\it e.g.}, the recent reviews \cite{Cloet:2013jya, Eichmann:2016yit,Huber:2018ned,Sanchis-Alepuz:2017jjd}  
as well as references therein. All expressions in the following are understood to be formulated in 
Euclidean momentum space, {\it i.e.}, after a Wick rotation. 

In the DSE/BSE formalism, the fully-dressed quark-photon vertex $\Gamma^{\mu}$, which describes the interaction bet\-ween quarks and photons in a quantum field theory, can be obtained as the solution of an inhomogeneous BSE
\begin{eqnarray}
\label{eq:inhomBSE_vector}
\bracket{\Gamma^{\mu}}_{a\alpha,b\beta}\bracket{p,Q}&=&
Z_2 \bracket{\gamma^\mu}_{ab} t_{\alpha\beta}\\
&+&\int_q K^{r\rho,s\sigma}_{a\alpha,b\beta}\bracket{Q,p,q} \, S_{r\rho,e\epsilon}\bracket{k_1}
\nonumber \\
&\times& \bracket{\Gamma^{i,\mu}}_{e\epsilon,n\nu}\bracket{Q,q}S_{n\nu,s\sigma}\bracket{k_2} \, .
\nonumber
\end{eqnarray}
Here, $Q$ is the photon momentum, $p$ is the relative momentum between quark and antiquark, $q$ is an internal relative momentum which is integrated over, the internal quark and antiquark momenta are defined as $k_1=q+Q/2$ and $k_2=q-Q/2$, respectively, such that $Q=k_1-k_2$ and $q=(k_1+k_2)/2$. Latin  letters represent Dirac indices, and Greek letters represent flavour indices. 
The isospin structure of the vertex is given by 
$t_{\alpha\beta}= \textrm{diag}\bracket{\nicefrac{2}{3},\nicefrac{-1}{3}}$. The Dirac structure of the vertex can be expanded in a basis consisting of twelve elements \cite{Miramontes:2019mco}, and all of them are considered in our calculation.

Similarly, mesons as bound states of two quarks are described in this framework by Bethe-Salpeter amplitudes $\Gamma$ which are obtained as solutions of a homogeneous BSE, 
\begin{dmath}
\label{eq:homogeneousBSE}
\bracket{\Gamma}_{a\alpha,b\beta}\bracket{p,P}= \int_q K^{r\rho,s\sigma}_{a\alpha,b\beta}\bracket{P,p,q} \times S_{r\rho,e\epsilon}\bracket{k_1}\bracket{\Gamma}_{e\epsilon,n\nu}\bracket{q,P}S_{n\nu,s\sigma}\bracket{k_2}~,
\end{dmath}
where for clarity we have here used $P$ for the total meson momentum (instead of $Q$ as above). For pions, the Dirac part of the Bethe-Salpeter amplitude $\Gamma$ can be expanded in a tensorial basis with four elements.

In the equations above, the interaction kernel $K$ describes the interaction between quark and antiquark, and $S$ is the fully-dressed quark propagator. We will discuss in detail the interaction kernels below. The quark propagator $S(p)$ is obtained as the solution of the quark DSE,
\begin{dmath}
S^{-1} = S_0^{-1} - Z_{1f} \int_q \gamma_\mu S(q) \Gamma^{qgl}_\nu(q,k) D_{\mu \nu}(k)~,
\label{eq:quarkDSE}
\end{dmath}
with $S_0^{-1}$ the renormalised bare propagator,
\begin{equation}
S_0^{-1}(p) = Z_2\left(i \Slash{p} + Z_m m\right)~,
\end{equation}
and $Z_{1f}$, $Z_2$ and $Z_m$ are renormalisation constants, $m$ is the (renormalisation-point dependent) current quark mass, $\Gamma^{qgl}$ is the full quark-gluon vertex and $D_{\mu \nu}$ is the full gluon propagator which, in the Landau gauge, is parametrised as
\begin{equation}
D_{\mu \nu}(k) = \left( \delta_{\mu \nu} - \frac{k_\mu k_\nu}{k^2}\right) \frac{Z(k^2)}{k^2}~,
\end{equation}
with $Z(p^2)$ being the gluon dressing function. For simplicity, we have suppressed the color indices. 

\begin{figure*}[t!]
\centerline{%
\includegraphics[width=17cm]{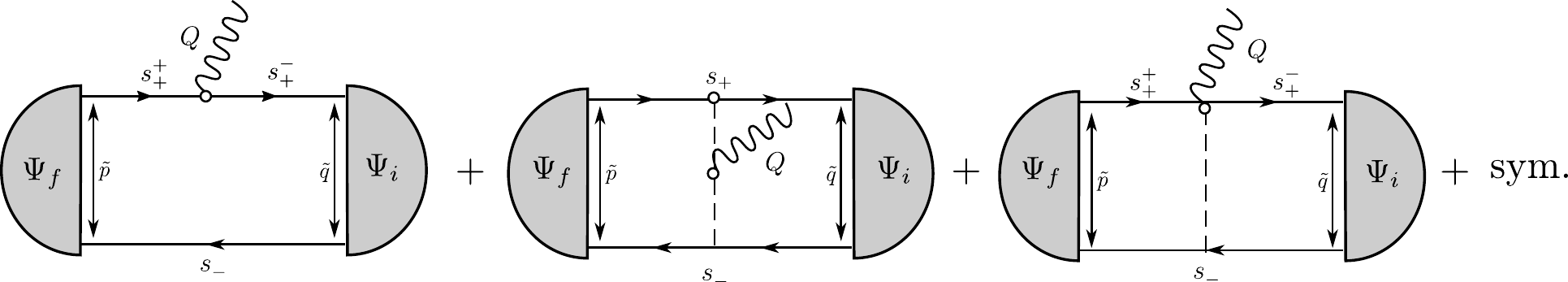}}
\caption{Diagrams relevant for the calculation of the pion form factor in the truncation employed herein, as determined by the gauging method. The impulse approximation implies considering the first diagram and the corresponding permutation only.} 
\label{Fig:BI_diagrams}     
\end{figure*}

\subsection{Interaction Kernels}

The interaction kernel $K$ in eqs.\ \eqref{eq:inhomBSE_vector} and \eqref{eq:homogeneousBSE}
encodes all possible interactions processes between a quark and an antiquark. In a diagrammatic representation, it contains a sum of infinitely many terms. In practical calculations, the expansion of the interaction kernel must be truncated to a sum of a finite number of terms, chosen such that the relevant dynamics and global symmetries are correctly implemented. Chiral symmetry and its dynamical breaking ensures that pions are  massless bound states in the chiral limit as a consequence of Goldstone’s theorem. On the other hand, U(1) vector  symmetry ensures charge conservation in electromagnetic processes. Chiral symmetry will be correctly implemented in the DSE/BSE formalism only if the kernel fulfills the axial-vector Ward-Takahashi identity (Ax-WTI)
\begin{dmath}
i\Sigma_{ar}\bracket{p_+}\gamma^5_{rb}t^i_{\alpha\beta}
+i\gamma^5_{ar}\Sigma_{rb}\bracket{p_-}t^i_{\alpha\beta}=\int_q K^{r\rho,s\sigma}_{a\alpha,b\beta}\bracket{Q,p,q}\left[i~t^i_{\rho\nu}\gamma^5_{rn}S_{n\nu,s\sigma}\bracket{q_-} +i~S_{r\rho,e\epsilon}\bracket{q_+}\gamma^5_{es}~t^i_{\epsilon\sigma}\right]~,\label{eq:AxWTI_BSEkernel}
\end{dmath}
with $\Sigma$ the quark self-energy and $v_\pm=v\pm Q/2$. Similarly, vector symmetry will be correctly implemented if the kernel satisfies the vector Ward-Takahashi identity (V-WTI) 
\begin{dmath}
i\Sigma_{ab}\bracket{p_+}t^i_{\alpha\beta}
-i\Sigma_{ab}\bracket{p_-}t^i_{\alpha\beta}=\int_q K^{r\rho,s\sigma}_{a\alpha,b\beta}\bracket{Q,p,q}\left[i~t^i_{\rho\nu}S_{r\nu,s\sigma}\bracket{q_-}-i~S_{r\rho,s\epsilon}\bracket{q_+}~t^i_{\epsilon\sigma}\right]~.\label{eq:VWTI_BSEkernel}
\end{dmath}

In DSE/BSE studies the most widely used truncation is the so-called rainbow-ladder (RL) truncation, whereby the BSE kernel consists of a vector-vector gluon exchange, namely (omitting again color indices)
\begin{dmath}
K^{r\rho,s\sigma}_{a\alpha,b\beta}\bracket{Q,p,q}=\alpha\bracket{k^2}\gamma^\mu_{ar}\gamma^\nu_{sb}D^{\mu\nu}\bracket{k}\delta^{\alpha\rho}\delta^{\sigma\beta}~,\label{eq:RLkernel}
\end{dmath}
with $k=p-q$ the gluon momentum. In order to preseve the Ax-WTI and V-WTI the kernel \eqref{eq:RLkernel} is used in combination with a truncated quark DSE, defined by the replacement 
\begin{equation}
Z_{1f} \gamma_{\mu} Z(k^2)\Gamma_{\nu}^{\text{qgl}}(q,p)  \rightarrow Z_2^2 \gamma_{\mu} 4\pi \alpha(k^2) \gamma_{\nu}
\end{equation}
such that $\alpha(k^2)$ provides an effective coupling that describes the strength of the quark-antiquark interaction. 
To parametrise this effective interaction we use the Maris-Tandy model \cite{Maris:1997tm, Maris:1999nt}
\begin{dmath}\label{eq:MTmodel}
\alpha(q^2)=
 \pi\eta^7\left(\frac{q^2}{\Lambda^2}\right)^2
e^{-\eta^2\frac{q^2}{\Lambda^2}}+\frac{2\pi\gamma_m
\big(1-e^{-q^2/\Lambda_{t}^2}\big)}{\textnormal{ln}[e^2-1+(1+q^2/\Lambda_{QCD}
^2)^2]}~,
\end{dmath}
where the second term on the right-hand side reproduces the one-loop QCD behavior of the quark propagator in the ultraviolet, and the  Gaussian term provides enough interaction strength for dynamical chiral symmetry breaking to take place. The model parameters $\Lambda$ and $\eta$ are determined as explained in the next section. The scale $\Lambda_t=1$~GeV is introduced for technical reasons and has no impact on the results.  For the anomalous dimension we use $\gamma_m=12/(11N_C-2N_f)=12/25$ with $N_f=4$ flavours and $N_c=3$ colours. For the QCD scale we use $\Lambda_{QCD}=0.234$ GeV.

In the RL truncation, bound states cannot develop a decay width, correspondingly their masses are real numbers. Note that bound states, determined as solutions of BSEs, appear as poles in Green's functions with the corresponding quantum numbers. In the RL approximation these poles occur for real 
(and in the convention employed here, negative) momentum-squared values in certain kinematic configurations. In particular, the photon being described by a vector field, electrically neutral vector mesons appear as poles of the quark-photon vertex. In the RL approximation these poles are located at  negative and real values of $Q^2$ (for which $M^2=-Q^2$ with $M$ being the mass of the vector meson). Such poles in the quark-photon vertex also manifest as poles in the calculation of time-like form factors, in contradiction with phenomenology. Generally speaking, any physical phenomenon that is triggered by the presence of virtual intermediate particles, as, {\it e.g.}, decays, will be absent from any calculation using the RL truncation only.

It is possible to improve the RL truncation in this respect by re-introducing\footnote{Note 
that effects of intermediate virtual states like the decay of hadrons are in principle  present in the full quark-gluon 
vertex which, however, is drastically simplified in the RL truncation.}
 the presence of intermediate particles explicitly. The simplest implementation of such an idea was introduced in \cite{Fischer:2007ze,Fischer:2008sp}\footnote{See, however, ref.\ \cite{Alkofer:1993gu} for considering pion loop 
contributions to the electromagnetic pion radius in the DSE/BSE approach. 
An alternative way is to introduce two-pion states via an explicit two-pion component in the bound state amplitude, 
see ref.\ \cite{Santowsky:2020pwd} and references therein.}, where, 
based on the role of the pion as lightest 
hadron, explicit pion-quark interactions were introduced in the truncated quark DSE and in the BSE kernel $K$, 
with the pion-quark interaction vertex given by the pion Bethe-Salpeter amplitude 
$\Gamma$, calculated consistently via a truncated BSE.
The corresponding additional BSE kernels are given in Appendix \ref{kernels} and shown in Fig.~\ref{fig:kernels}, and the technical difficulties arising for time-like momenta when those kernels are used in BSEs have been described in detail in \cite{Miramontes:2019mco}. This type of kernels enable the possibility of intermediate virtual decays in the BSE interaction kernel to occur. As a consequence, certain BSE solutions signal a finite decay width and thus represent (a) hadron resonance(s). {\it E.g.}, the description of the $\rho$-meson as a finite-width resonance is then mostly due to the intermediate process $\rho \rightarrow \pi\pi$ (see \cite{Williams:2018adr} for a treatment in the here discussed  approach as well as \cite{Jarecke:2002xd,Mader:2011zf} and references therein for respective calculations based on DSEs and
BSEs),
the partial decay width to the latter process representing more than 99\% \ of the total $\rho$  decay width.
 Additionally, and thereby completing the physical effects of intermediate virtual $\pi\pi$-states, in this truncation the quark-photon vertex develops a multi-particle branch cut along the negative real $Q^2$ axis, starting as expected at the two-pion production threshold \cite{Miramontes:2019mco}.
Clearly, including these two effects of the intermediate virtual $\pi\pi$-states is especially important  when it comes to the calculation of the pion form factor for time-like momenta in the sub-GeV kinematic region.

We wish to stress here that, for computational feasibility, for the pion vertices in \eqref{eq:BSEkernel_tchannel}--\eqref{eq:BSEkernel_uchannel} we used the leading $\gamma_5$ component of the pion Bethe-Salpeter amplitude in the chiral limit, given by $B/f_\pi$ with $B$ one of the quark's dressing functions, see Eq.~\eqref{eq:pion_chiral}. On the other hand, the pion amplitudes used in the form factor calculation of Eq.~\eqref{eq:Current} are considered in full, including their leading and sub-leading contributions. In that regard, our calculations contain \textit{two types of treatments of pions}.

Even though the ``pionic'' kernels \eqref{eq:BSEkernel_tchannel}--\eqref{eq:BSEkernel_uchannel} are phenomenologically justified and, as we will see in the next section, constitute a first step in the correct direction, it must be noted that they have not been (yet) rigorously derived from QCD. Lacking a solid quantum-field theoretical basis the use of these kernels comes with some shortcomings, especially it implies that the Ax-WTI and V-WTI are not fulfilled simultaneously. Indeed, one can choose to preserve either the Ax-WTI (and hence chiral symmetry) or the V-WTI (and hence charge conservation), but not both \cite{Fischer:2007ze,Miramontes:2019mco}. It turns out, however, that the respective violation of either of them induce typically  only small errors  in physical observables, as we will demonstrate for some quantities in the next section.

\subsection{Form factor calculation}

Meson form factors are extracted from a current $J^{\mu}$ encoding the coupling of a meson to an external electromagnetic current. In the BSE framework, the current is calculated by means of the coupling of an external photon to each of the constituents of the bound state, as specified by a procedure known as \textit{gauging} and developed in \cite{Haberzettl:1997jg,Kvinikhidze:1998xn,Kvinikhidze:1999xp,Oettel:1999gc,Oettel:2000jj}. The conserved current $J^{\mu}$ that describes the coupling of a single photon with a quark-antiquark, a three-quark or other multi-quark system is given by,
\begin{equation}
J^{\mu} = \bar{\Psi}_f G_0 (\mathbf{\Gamma}^{\mu} - K^{\mu}) G_0 \Psi_i ~,
\label{eq:Current}
\end{equation} 
with $\Psi_{i,f}$ the incoming and outgoing Bethe-Salpeter amplitudes of the meson, the baryon or some other multi-quark state, and $G_0$ represents the appropriate product of dressed quark propagators. This equation is shown diagrammatically for the  quark-antiquark--meson case in Fig.~\ref{Fig:BI_diagrams}. The term $\Gamma^{\mu}$ represents the impulse approximation diagrams where the photon couples to the valence quarks only
\begin{equation}
\mathbf{\Gamma}^{\mu} = \left(S^{-1} \otimes S^{-1} \right)^{\mu} = \Gamma^{\mu}\otimes S^{-1} + S^{-1}\otimes \Gamma^{\mu}~,
\end{equation}
with $\Gamma^{\mu}$ the quark-photon vertex. The term $K^{\mu}$ describes the interaction of the photon with the 
Bethe-Salpeter kernel, which in our truncation includes the coupling of the photon to the quark-pion vertex and to the 
propagating pions 
(second and third diagram in Fig.\ \ref{Fig:BI_diagrams}).
 Including both terms in \eqref{eq:Current} is necessary in order to implement current conservation precisely. 
Note that the s- and u-channel pion decay terms (most right diagram in Fig.\ \ref{fig:kernels}) do not contribute to 
the term  $K^{\mu}$, also not via seagulls, the reason being that trying to include them would leave a $\pi$ $\pi$ $\pi$ 
amplitude on one side of  the respective full diagram.
However, the new vertices appearing in the term $K^{\mu}$, 
the coupling of the photon to the  quark-pion vertex and to the propagating pions,
 represent an enormous computational challenge. 
Given the exploratory purpose of the present calculation, we thus decided to omit the coupling of the photon to the quark-pion vertex and to the propagating pions, and to consider the impulse diagram only (first diagram in Fig.~\ref{Fig:BI_diagrams}). As we will show in the next section, the thereby implied violation of charge conservation is at the level of approximately one percent.

\section{\label{sec:results} Results}

Following the formalism sketched above, we have calculated the pion electromagnetic form factor in the time-like $Q^2<0$ domain. For comparison with previous calculations we will also show results in the space-like $Q^2>0$ domain.

In our discussion in the preceding section, it remained to be explained how the parameters $\eta$ and $\Lambda$ of our interaction model \eqref{eq:MTmodel} as well as the value of the current quark mass are fixed. It is customary in studies using the RL truncation to adjust those parameters such that the pion decay constant agrees with the experimental value\footnote{Note that, for this observable, the result is quite independent of the value of $\eta$ around $\eta=1.8$ and thus, effectively, only the  parameter $\Lambda$ has to be adjusted.}. 
In ref.\ \cite{Miramontes:2019mco} the quark-photon vertex has been calculated for the first time with the above discussed interaction kernels taken into account and also used herein. The parameters, including the isospin symmetric light quark current mass $m_q$, were adjusted such that the pion mass and decay constant as well as the $\rho$-meson mass have been correctly reproduced in the employed approximation. Here, only the gluon- and pion-exchange kernels need to be used for fixing the parameters because the pion decay kernels \eqref{eq:BSEkernel_schannel} and \eqref{eq:BSEkernel_uchannel} do not contribute to the pion BSE. 

For the present exploratory calculation, we choose to adjust the parameters in a slightly different and simpler manner, especially as we aim at a qualitative understanding of the physical mechanisms involved in determining the shape of the pion form factor, and not so much at achieving an accurate quantitative agreement with experiment. 
First, we set initially $\eta=1.5$. Second, although we assume (confirmed by our calculation) that the pion-decay kernels will not only move the $\rho$-meson pole into the complex plane but also shift down its real value, we nevertheless adjust the parameter $\Lambda$ such that we obtain a $\rho$-meson mass close to the phenomenological value already in the calculations with gluon- and pion-exchange kernels only. Third, we require to reproduce a quite accurate value for the pion decay constant.
Finally, we adjust $m_q$ to obtain the correct value for the pion mass as well. In this way we chose the model parameters to be $\eta=1.5$, $\Lambda=0.78$~GeV and $m_q=6.8$~MeV at a renormalisation scale $\mu=19$~GeV. As a rudimentary test of model dependence we additionally perform the calculations for $\eta=1.6$ as well (keeping $\Lambda$ and $m_q$ unchanged). 
The results for the pion mass and decay constant as well as for the $\rho$-meson and $\omega$-meson masses, without the pion decay kernels being taken into account, are shown in Tab.~\ref{tab:masses}.

\begin{table}[t]\caption{\label{tab:masses}
The pion mass $m_\pi$, the pion decay constant $f_\pi$, the $\rho$-meson and $\omega$-meson masses $m_\rho$ and $m_\omega$ for the two different parameterizations of the model used herein
and for the case with rainbow-ladder and pion-exchange kernels but without decay kernels are shown. 
The light quark mass has been set to $m_q=0.0068~$GeV.
The rightmost column displays the extracted $\rho$-meson pole position defined as 
$M_{pole}^2=M_\rho^2-iM_\rho \Gamma_\rho $, as discussed in the text. 
All values for dimensionful quantities are given in GeV.}
\begin{ruledtabular}
\begin{tabular}{l|cccc||cc}
$\Lambda=0.78$&
$m_\pi$&
$f_\pi$&
$m_\rho$&
$m_\omega$&
$M_\rho$&
$\Gamma_\rho$\\
\colrule
$\eta=1.5$ & 0.139 & 0.138 & 0.768 & 0.778  & 0.750 & 0.100\\
$\eta=1.6$ & 0.126 & 0.138 & 0.774 & 0.784  & 0.759 & 0.105\\
\end{tabular}
\end{ruledtabular}
\end{table}

As mentioned in the previous section, the full QCD quark-photon vertex possesses poles reflecting 
the masses and widths of the electrically neutral vector meson resonances.
Correspondingly, and as discussed in detail in refs.\ \cite{Miramontes:2019mco,Williams:2018adr}, a solution for the quark-photon vertex in the DSE/BSE framework allows to extract the $\rho$-meson mass and width via the position of the poles of the vertex dressing functions. For the RL truncation 
as well as for the RL plus pion exchange approximation this pole is located on the real negative $Q^2$ axis indicating that the $\rho$ mesons were stable for those truncations. Including the $s$- and $u$-channel decay kernels \eqref{eq:BSEkernel_schannel} and \eqref{eq:BSEkernel_uchannel}, the pole of the dressing functions moves into the complex plane and can be extracted from the data on the real axis via a Pad\'e fit. Parametrising the pole position as 
$M_{pole}^2=M^2_\rho -iM_\rho \Gamma_\rho $, we extract the corresponding results for  the 
$\rho$-meson mass and width, $ M_\rho$ and $\Gamma_\rho $ for this truncation (see Tab.~\ref{tab:masses}) in reasonable agreement with the experimental values. Hereby, it has to be noted that underestimating the $\rho$-meson width does not 
come unexpected because taking into account only the leading $\gamma_5$ component of the pion Bethe-Salpeter amplitude
 in the kernels \eqref{eq:BSEkernel_tchannel}--\eqref{eq:BSEkernel_uchannel} misses some strengths therein. Whether 
considering in addition also the sub-leading pion amplitudes will provide a much better result for the $\rho$-meson width 
can only be answered by performing the corresponding calculation. This computation is then, however, an order of magnitude 
more expensive than the present exploratory calculation.

It is interesting to note that the pion exchange kernels lift the degeneracy in between the isovector 
$\rho$- and the isosinglet $\omega$-meson present at the level of RL calculations the interaction kernels of which are flavour blind and thus flavour U(2) (resp., flavour U$(N_f)$) symmetric. The pion 
exchange kernels lead to a splitting such that $m_\omega - m_\rho = 10$ MeV which compares favorably with the experimental splitting of 7 - 8 MeV. 

We turn now to the calculation of the pion form factor. 

\begin{figure}[ht]
\includegraphics[width=0.49\textwidth]{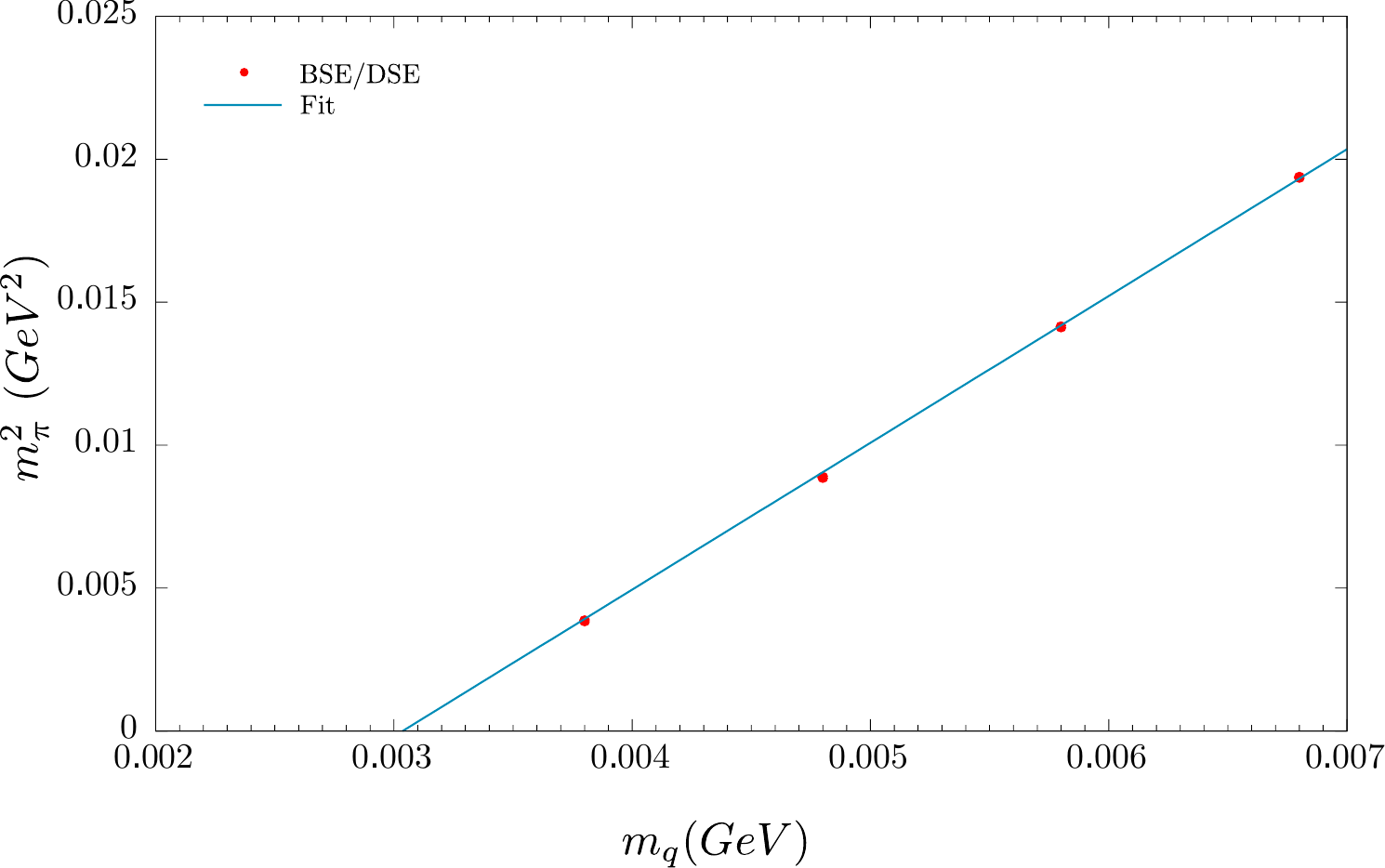}
\caption{The pion mass squared $m_\pi^2$ vs.\ the current mass $m_q$($\mu=19$~GeV) for the employed truncation.}
\label{fig:mpiFif}     
\end{figure}

As already indicated in the previous section, there are, besides restricting to the leading pion amplitude in the 
kernels\eqref{eq:BSEkernel_tchannel}--\eqref{eq:BSEkernel_uchannel},
  two major approximations that we must perform in order to keep the calculation technically manageable. First, using the 
decay kernels as described in this work entails that one must choose whether the axial-vector or the vector WTI are 
preserved while the other one is violated. Following \cite{Miramontes:2019mco} we choose to preserve the vector identity. 
A violation of the axial-vector WTI is manifested, among others, in the pion not being massless in the chiral limit, and 
therefore the value of the current mass for which the pion becomes massless allows for a quantification of the violation 
of the axial-vector WTI. In Fig.~\ref{fig:mpiFif} we therefore show the evolution of the pion mass with varying $m_q$ in the 
employed truncation. First, the relation is linear as expected from the Gell-Mann--Oakes--Renner relation which
is a direct consequence of the dynamical breaking of chiral symmetry. Second, the pion does not 
become massless in the chiral limit but for a value of the current mass 
$m_q^{(0)}$($\mu=19$~GeV) = 3 MeV. On the one hand, this explains the relatively large value of 
$m_q$($\mu=19$~GeV) = 6.8 MeV we needed to obtain the correct pion mass: The related 
explicitly chiral-symmetry-breaking term is $m_q - m_q^{(0)}$ = 3.8 MeV, and thus much closer to what is expected from the known parameters of QCD. Second, as the masses of the vector mesons
depend linearly on the current mass the induced error on $m_\rho$ and $m_\omega$ is of the order
of $m_q^{(0)}$ = 3 MeV, and thus it is as small or even smaller than other uncertainties in our calculation of the vector meson masses.

\begin{figure}[hb]
\includegraphics[width=0.49\textwidth]{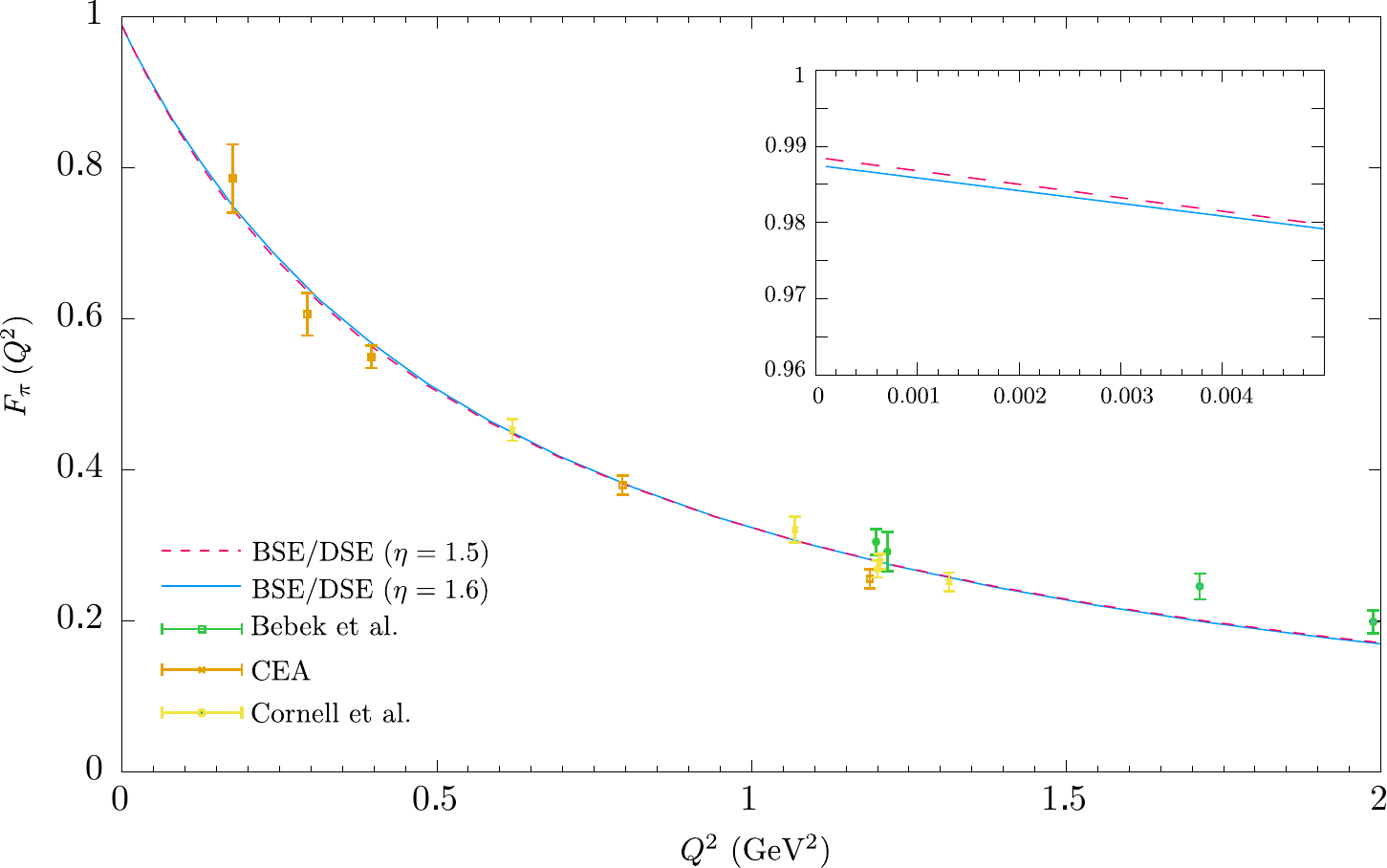}
\caption{Pion form factor in the spacelike $Q^2>0$ domain for the model parameters $\eta=1.5$ and $\eta=1.6$ as described in the text and compared to experimental data.}
\label{fig:FF_spacelike}    
\end{figure}

Second, in the calculation of the form factor we use the impulse approximation which, in this context, implies neglecting the second and third diagrams in Fig.~\ref{Fig:BI_diagrams}. The consequence of discarding diagrams is the violation of charge conservation or, equivalently, a deviation from $F_\pi(0)=1$. As can be seen in the inset in Fig.~\ref{fig:FF_spacelike}, this effect is of the order of $\sim 1\%$ only.

\begin{figure*}[ht]
\centerline{%
\includegraphics[width=0.8\textwidth]{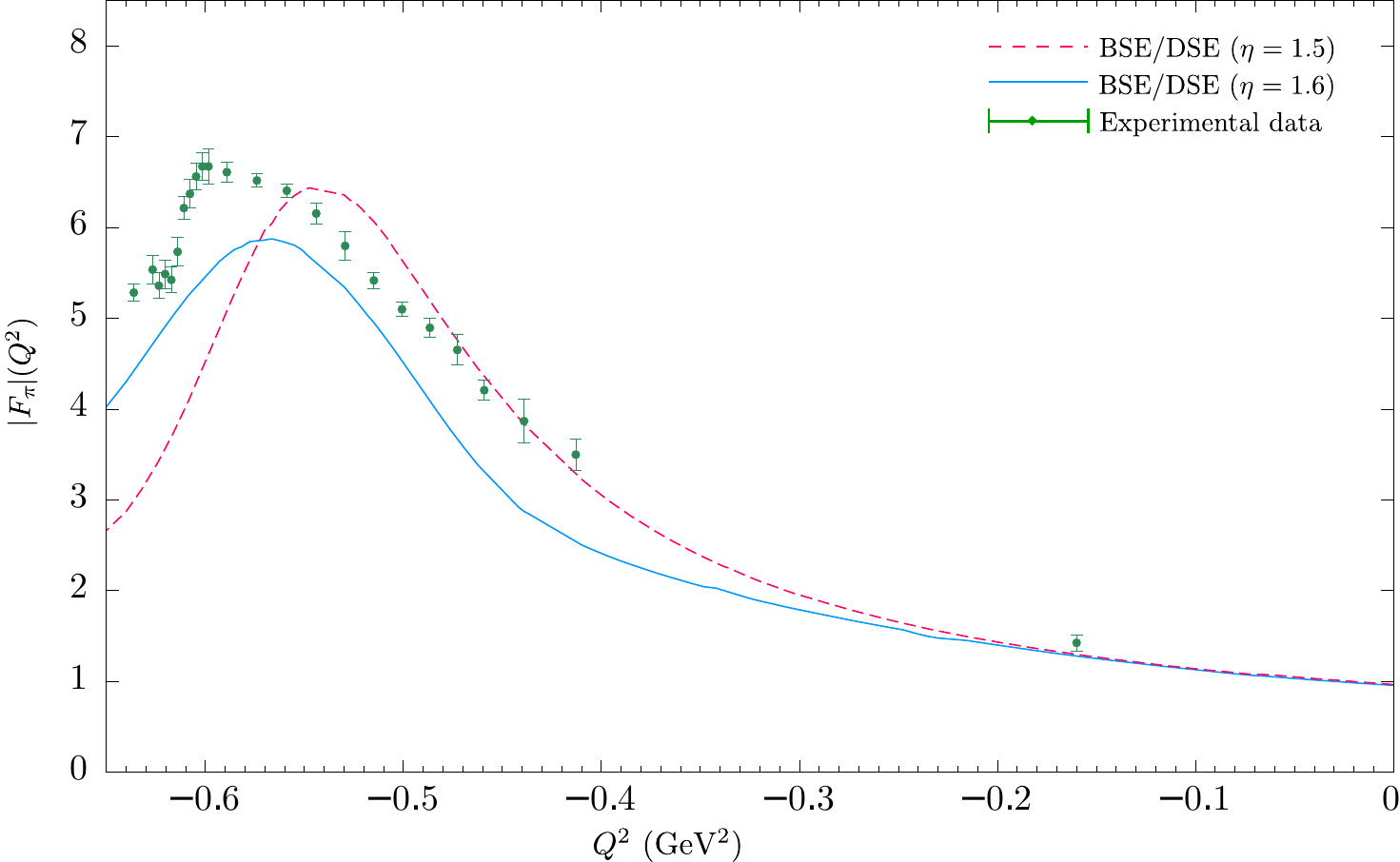}}
\caption{Absolute value of the pion form factor in the time-like $Q^2<0$ domain for  the model parameters $\eta=1.5$ and $\eta=1.6$ as described in the text and compared to experimental data.}
\label{fig:FF_abs_timelike}    
\end{figure*}

As can be also seen from Fig.~\ref{fig:FF_spacelike}, the results for the pion form factor in the space-like $Q^2>0$ regime are practically independent from the value of the $\eta$ parameter of the model.
Even more remarkable is the fact that our calculation shows a very good agreement with experimental data in the space-like domain even though, as evident from the above discussion, we aimed at including all physical effects which are important in the time-like regime. An interpretation of this result in view of the 
dispersion relation \eqref{eq:DispRel} provides an indication that the imaginary part in the time-like region is 
precisely enough reproduced to provide very good results for the space-like form factor.

\begin{figure*}[ht]
\centerline{%
\includegraphics[width=0.8\textwidth]{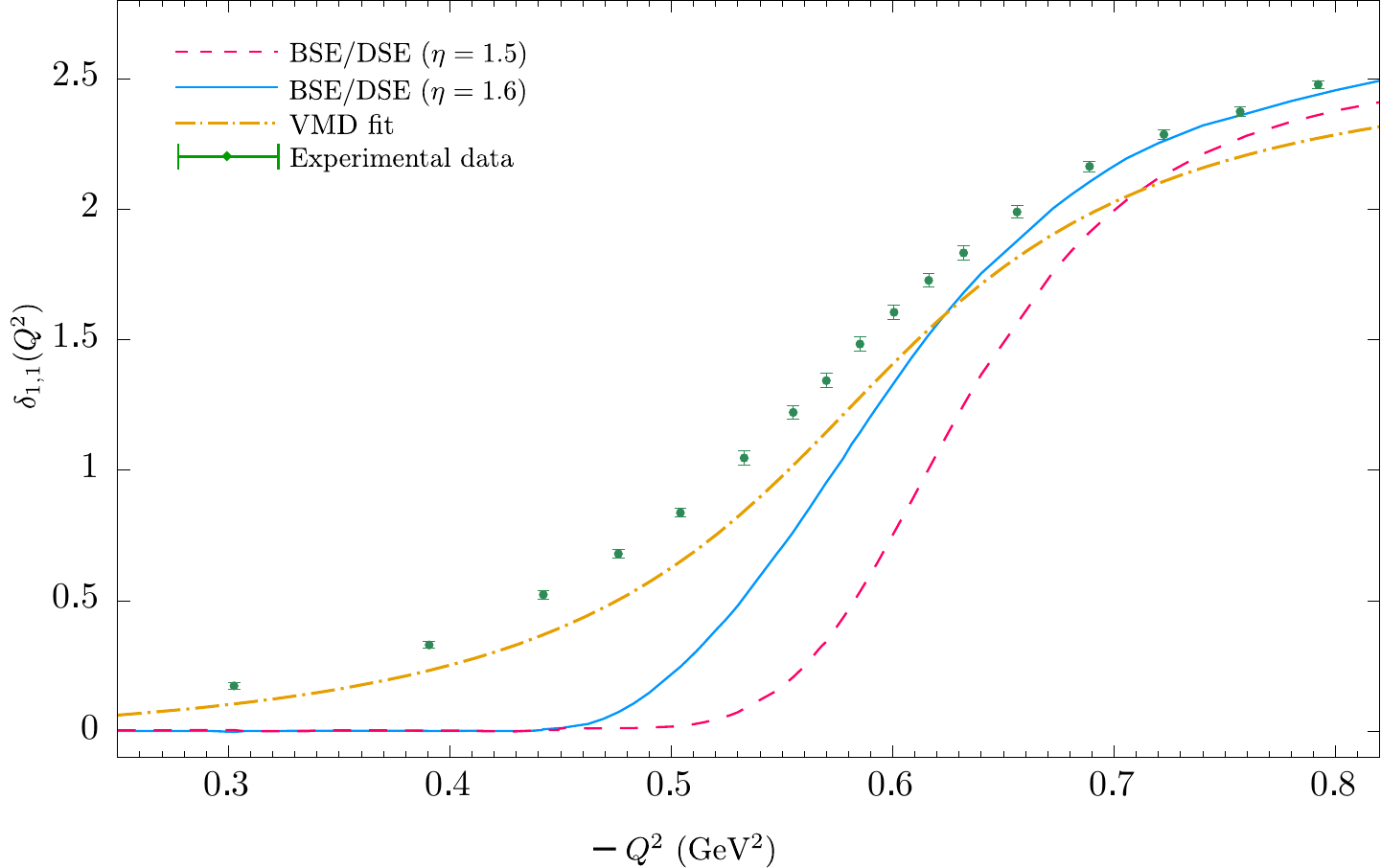}}
\caption{Phase of the pion form factor in the time-like $Q^2<0$ domain for the model parameters $\eta=1.5$ and $\eta=1.6$ as described in the text and compared to experimental data on pion-pion phase shift.}
\label{fig:FF_phase_timelike}    
\end{figure*}

We show our results for the pion form factor for time-like ($Q^2<0$) virtualities in Figs.~\ref{fig:FF_abs_timelike} and \ref{fig:FF_phase_timelike}. As discussed in the previous section, as a consequence of the decay kernels in our truncation, the pion form factor develops a branch cut along the real negative axis starting from the two-pion threshold $Q^2=-4m_\pi^2$, induced by the corresponding cut in the quark-photon vertex \cite{Miramontes:2019mco}. Hence, in that region the (complex) form factor is defined from its analytic continuation as $F(Q^2+i\epsilon)$. In the numerical calculations we have typically chosen  
$\epsilon=0.0001$~GeV$^2$ after verifying that this value is small enough to not disturb the presented results. In Fig.~\ref{fig:FF_abs_timelike} we present the absolute value for the model parameter $\eta=1.5$ and $\eta=1.6$, with the remaining model parameters kept constant, as discussed above. As a manifestation of the fact that the $\rho$-meson pole in the quark-photon vertex moves into the complex plane when the decay kernels are included in the calculation, the pion form factor develops a bump on the real and negative $Q^2$ axis with an approximately correct height and width. Therefore our calculation overcomes a major
deficiency of the RL truncation, without or even with the pion exchange term, for which the form factor diverges instead at the $Q^2$-value corresponding to the $\rho$-meson mass in those truncations,
see e.g.~\cite{Maris:1999bh,Krassnigg:2004if,Eichmann:2019tjk}. 
This constitutes already one main result of the 
here presented investigation. 

We note, however, that, contrary to the results for space-like regime, the position and height of the bump of the form factor depends strongly on the value of the $\eta$ parameter shape of the form factor. Of course,  this reflects the different positions of the $\rho$-meson pole, {\it cf.} Tab.~\ref{tab:masses}. Nevertheless, there are features that appear to be independent of $\eta$, most prominently that the height of the bump is underestimated. As expected from the discussion in Sec.~\ref{sec:PionFF} the form factor behaves smoothly after the bump, in contrast to the sharp dip in the experimental data. Even though an unambiguous analysis of the origin of such discrepancies can only result from the inclusion of all relevant physical mechanisms in our calculations, it is evident from the analysis performed in  Sec.~\ref{sec:PionFF}
that one of the main missing elements is the $\rho$-$\omega$ mixing due to isospin breaking, 
which is completely absent in the present study.

Particularly sensitive to the deficiencies of our truncation is the phase of the form factor, shown in Fig.~\ref{fig:FF_phase_timelike}. Even though our data shows the expected behaviour near the resonance value, it severely underestimates the experimental data, particularly in the elastic region. This is a manifestation of the absence of some hadronic effects in our approximation scheme, which only includes those stemming from the resonance complex pole in the quark-photon vertex and the $\rho\rightarrow\pi\pi$ induced branch cut. In addition to the isospin breaking effects discussed above, and which would be more relevant in the region above the resonance, the impulse approximation used in our calculation of the form factor entails that effects coming from the coupling of the photon to the intermediate hadrons, via its coupling to the exchanged pion or to the quark-pion vertex (see Fig.~\ref{Fig:BI_diagrams}), are missing. It has been shown \cite{Cotanch:2002vj} that considering only impulse-like diagrams leads to a very small pion-pion scattering amplitude in the elastic region in the isospin $I=1$ channel. This, due to unitarity, implies a very small value of the imaginary part of the pion form factor in the elastic region (which is, in fact, what we observe) which translates into a very small phase, as seen in Fig.~\ref{fig:FF_phase_timelike} (and as could be inferred from Watson's theorem).

Last but not least, we are comparing Pad\'e fits, resp.\ rational fits of the order (3,3), for the real part and 
for the imaginary part of the form factor to the expression \eqref{eq:VMDsimplified}. Trying first,
\begin{eqnarray}
Re \, F_\pi (Q^2) - F_\pi (0) & \approx &
- \frac {a_0 + a_1 Q^2 + a_2 (Q^2)^2 + a_3 (Q^2)^3}{b_0 + b_1 Q^2 + b_2 (Q^2)^2 + b_3 (Q^2)^3} \nonumber \\ 
Im \, F_\pi (Q^2) & \approx & \frac {c_0 + c_1 Q^2 + c_2 (Q^2)^2 + c_3 (Q^2)^3}
{d_0 + d_1 Q^2 + d_2 (Q^2)^2 + d_3 (Q^2)^3} \, , \nonumber \\
\end{eqnarray}
we obtain tiny values for the coefficients $a_0$, $a_3$, $b_3$, $c_0$, $c_3$ and $d_3$. Note that this {\em confirms the 
structure expected from the VMD form} \eqref{eq:VMDsimplified}. We repeated the fits for
\begin{eqnarray}
Re \, F_\pi (Q^2) - F_\pi (0) & \approx &
- \frac {a_1 Q^2 + a_2 (Q^2)^2 }{b_0 + b_1 Q^2 + b_2 (Q^2)^2} \nonumber \\
Im \, F_\pi (Q^2) & \approx & \frac { c_1 Q^2 + c_2 (Q^2)^2 }
{d_0 + d_1 Q^2 + d_2 (Q^2)^2 } \, .
\end{eqnarray}
The coefficients resulting from these fits as well as the ones resulting from expression \eqref{eq:VMDsimplified} 
are given in table~\ref{tab:Coeffs}. From this we conclude that the expression based on the VMD is an
astonishingly good representation of our results. Therefore, our investigation makes it plausible that the VMD picture 
can be derived from QCD. At least, the results of the here presented microscopic approach give a strong hint into this 
direction.

\begin{table}[h]\caption{\label{tab:Coeffs}
The coefficients of the rational fits to the pion form factor as discussed in the text. 
All values for dimensionful quantities are given in GeV.}
\begin{ruledtabular}
\begin{tabular}{l|ll||l}
& $\eta$ =1.5 & $\eta$ =1.6 & Eq.\ \eqref{eq:VMDsimplified} \\
\hline \hline
$a_1$ & 0.5587 & 0.4149 & 0.72\\
$a_2$ & 0.8828 & 0.6827 & 1.2 \\
\hline
$b_0$ & 0.3600 & 0.3600 & 0.36\\
$b_1$ & 1.2307 & 1.2517 & 1.2\\
$b_2$ & 1.0722 & 1.1000 & 1.037\\
\hline \hline
$c_1$ & 0.0591 & 0.0997 & 0\\
$c_2$ & 0.1295 & 0.2383 & 0.2308\\
\hline
$d_0$ & 0.3600 & 0.3600 & 0.36\\
$d_1$ & 1.1924 & 1.2464 & 1.2\\
$d_2$ & 0.9973 & 1.0916 & 1.0037\\
\end{tabular}
\end{ruledtabular}
\end{table}

\section{\label{sec:outlook} Conclusions and outlook}

In this work we have presented an exploratory study of the pion form factor in the DSE/BSE approach. Our focus has been
to explore how the interplay between hadron structure (as described by form factors) and the hadron spectrum (as described 
by resonance masses and widths) can be realised in the microscopic approach presented herein. In particular, we focused 
on the effect of intermediate pions in the BSE interaction kernel, the inclusion of which is sufficient to describe the $\rho$ meson as a resonance \cite{Williams:2018adr}. As elucidated by the detailed analysis in the last section our calculation represents 
a verification of the vector meson dominance picture and provides an explanation how at the quark level  vector meson dominance becomes effective.  A more complete calculation than the one presented here might then actually provide a 
derivation of vector meson dominance from QCD.

Despite the fairly drastic approximations used in this preliminary study, the agreement with experiment is remarkable. On the space-like side our calculations agree with experimental data at the quantitative level. For time-like momentum transfers 
the agreement is mostly qualitative and consistent with the fact that in our approximation scheme time-like physics is 
dominated by the lowest-lying $\rho$-meson resonance only. The absolute value of the calculated form factor features a 
bump at approximately the correct $Q^2$ region as caused by a resonance pole. Our result lacks, however, other features 
such as those caused by the isospin-breaking $\rho$-$\omega$ mixing and interference. The phase of the form factor also shows deficiencies caused by the employed impulse approximation. However, the overall qualitatively correct behaviour is, nevertheless, very encouraging for future studies on time-like phenomenology with BSE methods as it shows that the 
necessary computational techniques are getting more and more under control, and that within a functional-method-based
bound-state approach to QCD direct calculations in the time-like regime are becoming feasible.

Among the different physical mechanisms absent in our calculation, the most relevant one appears to be isospin breaking 
by the light quarks' masses and electric charges, and the different phenomena associated with it. 
The presented exploratory calculation paved the way to include in
a bound-state approach formulated in QCD degrees of freedom the effects of isospin violation, and hereby most 
prominently $\rho$-$\omega$ mixing, on the  time-like pion electromagnetic form factor. Thus including  isospin violation 
in a BSE approach is the topic of ongoing work. 
A further related topic is the study of the form factor for the coupling of a photon to three pions. On the one hand, this 
process is of special theoretical interest because the related form factor is at the soft point completely determined 
by the Abelian chiral anomaly. On the other hand, data of the COMPASS experiment are currently analysed \cite{Dominik},
and therefore experimental data for this form factor in the time-like region will become available. Combing previous studies
in the DSE/BSE approach for the space-like $\gamma \pi \pi \pi $ form factor 
\cite{Alkofer:1995jx,Bistrovic:1999dy,Cotanch:2003xv,Benic:2011rk,Eichmann:2011ec} 
with the techniques of the here presented calculation will thus enable a respective investigation of this form factor.

Another aspect we want to investigate is how to realise the idea of decay kernels like \eqref{eq:BSEkernel_schannel} and \eqref{eq:BSEkernel_uchannel} in a baryon bound state equation. This is a necessary step in order to tackle
 time-like nucleon form factors in the DSE/BSE approach, which is one of our major goals due to the increased effort 
and interest from the experimental side in highly precise measurements over a wide kinematical domain of the 
nucleon form factors.

\section*{Acknowledgements}

We thank Gernot Eichmann and Christian Fischer for a critical reading of the manuscript and helpful discussions.

This work was partially supported by the the Austrian Science Fund (FWF) under project number P29216-N36.\\
 A.S.\ Miramontes acknowledges CONACyT for financial support.

The numerical computations have been performed at the high-performance compute cluster of the University of Graz.

\appendix
\section{Interaction kernels}
\label{kernels}

\begin{figure*}[ht]
\centerline{%
\includegraphics[width=0.85\textwidth]{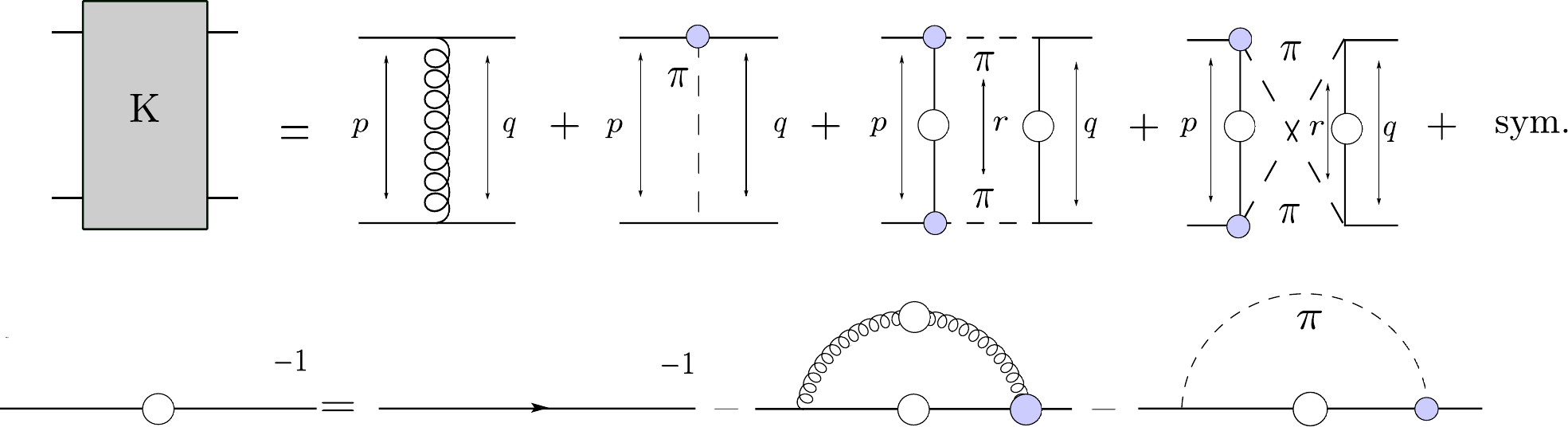}}
\caption{Truncations used herein for the BSE interaction kernel $K$ (upper diagram) and the quark DSE one (lower diagram). 
In the upper diagram, the terms on the right-hand side correspond to the rainbow-ladder, pion exchange, and s- and u-channel pion decay contributions to the truncation, respectively. The s- and u-channel pion decay terms do not contribute to the 
quark DSE.}
\label{fig:kernels}     
\end{figure*}

The BSE kernel representing the exchange of an explicit pionic degrees of freedom as defined in
\cite{Fischer:2007ze, Fischer:2008sp} reads
\begin{flalign}
&K^{(t)~ut}_{rs}(q,p;P) =\nonumber\\
&~~~~~~~~~~~\frac{C}{4} [\Gamma_{\pi}^j]_{ru} \left(\frac{p + q - P}{2}; p - q \right) [Z_2 \gamma^5]_{ts} D_{\pi}(p - q) \nonumber \\ \nonumber
 &~~~~~~~~+\frac{C}{4} [\Gamma_{\pi}^j]_{ru} \left(\frac{p + q - P}{2}; q - p \right) [Z_2 \gamma^5]_{ts} D_{\pi}(p - q) \\ \nonumber
 &~~~~~~~~+\frac{C}{4} [Z_2 \gamma^5]_{ru} [\Gamma_{\pi}^j]_{ts} \left(\frac{p + q + P}{2}; p - q \right) D_{\pi}(p - q) \\ 
 &~~~~~~~~+\frac{C}{4} [Z_2 \gamma^5]_{ru} [\Gamma_{\pi}^j]_{ts} \left(\frac{p + q + P}{2}; q - p \right) D_{\pi}(p - q)~,\label{eq:BSEkernel_tchannel}
\end{flalign}
\noindent in combination with the following truncation of the quark DSE
\begin{eqnarray}
S^{-1}(p) &=& S^{-1}(p)^{RL} - \frac{3}{2} \int_q \Bigg[Z_2 \gamma_5 S(q) \Gamma_{\pi}\left(\frac{p+q}{2}, q-p\right) \nonumber \\
&&  + Z_2 \gamma_5S(q)\Gamma_{\pi}\left(\frac{p+q}{2}, p-q\right)\Bigg] \frac{D_{\pi}(k)}{2}~,\label{eq:quarkDSE_tchannel}
\end{eqnarray}
with $S^{-1}(p)^{RL}$ being the right-hand-side of the quark DSE in the RL truncation with the gluon-mediated interaction as described in Sect.\ \ref{sec:DSE_BSE}. 
In Eqs.~\eqref{eq:BSEkernel_tchannel} and \eqref{eq:quarkDSE_tchannel} the pion propagator is taken as 
$D_{\pi}(k) = (k^2 + m_{\pi}^2)^{-1}$. The factor $3/2$ in Eq.~\eqref{eq:quarkDSE_tchannel} originates from the flavour traces, and in the same way the factor $C$ in \eqref{eq:BSEkernel_tchannel} should be obtained. When done in the quark-photon vertex, the flavour factor leads to $C=+3/2$. However, such value of $C$, violates the Ax-WTI while preserving the V-WTI. Since the V-WTI is related to charge conservation in electromagnetic form factors, we use $C=+3/2$ at the expense of violating the Ax-WTI.
Herein, the quark-pion vertex $\Gamma_\pi$ is taken to be the pion Bethe-Salpeter amplitude.

The pions in the kernel can also appear in the s- and u- channels \cite{Fischer:2007ze}. We used here a version of the kernels 
slightly different to the one in \cite{Fischer:2007ze} in order to be consistent with the construction for the t-channel, where one
 of the pion vertices is kept bare and the kernel is then symmetrised. They read

\begin{widetext}
\flaligne{
K^{(s)~he}_{da}(q,p,r;P)=~& \frac{C}{2}~D_{\pi}\left(\frac{P + r}{2}\right) D_{\pi}\left(\frac{P - r}{2}\right)~\left[[Z_2\gamma_5]_{dc} S_{cb}\left(p - \frac{r}{2}\right)[Z_2\gamma_5]_{ba}\right.\nonumber\\
&~~~~\times[\Gamma_{\pi}^j]_{hg} \left(q - \frac{P}{4} - \frac{r}{4}; \frac{r - P}{2} \right) S_{gf}\left(q - \frac{r}{2}\right)[\Gamma_{\pi}^j]_{fe} \left(q + \frac{P}{4} - \frac{r}{4}; -\frac{P + r}{2} \right)\nonumber\\
~&+~[\Gamma_{\pi}^j]_{dc} \left(p + \frac{P}{4} - \frac{r}{4}; \frac{P + r}{2} \right) S_{cb}\left(p - \frac{r}{2}\right)[\Gamma_{\pi}^j]_{ba} \left(p - \frac{P}{4} - \frac{r}{4}; \frac{P - r}{2} \right)\nonumber\\
&~~~~\left.\times[Z_2\gamma_5]_{hg} S_{gf}\left(q - \frac{r}{2}\right)[Z_2\gamma_5]_{fe}\right]~, 
\label{eq:BSEkernel_schannel}}

and

\flaligne{
K^{(u)~he}_{da}(q,p,r;P)=~& \frac{C}{2}~D_{\pi}\left(\frac{P + r}{2}\right) D_{\pi}\left(\frac{P - r}{2}\right)~\left[ [Z_2\gamma_5]_{dc}  S_{cb}\left(p + \frac{r}{2}\right) [Z_2\gamma_5]_{ba}\right.   \nonumber \\
 &~~~~\times [\Gamma_{\pi}^j]_{hg} \left(q - \frac{P}{4} - \frac{r}{4}; \frac{r - P}{2} \right) S_{gf}\left(q - \frac{r}{2}\right) [\Gamma_{\pi}^j]_{fe} \left(q + \frac{P}{4} - \frac{r}{4}; -\frac{P + r}{2} \right)\nonumber\\
~&+~[\Gamma_{\pi}^j]_{dc} \left(p + \frac{P}{4} + \frac{r}{4}; \frac{P - r}{2} \right) S_{cb}\left(p + \frac{r}{2}\right)[\Gamma_{\pi}^j]_{ba} \left(p - \frac{P}{4} + \frac{r}{4}; \frac{P + r}{2} \right)  \nonumber \\
 &~~~~\left.\times[Z_2\gamma_5]_{hg} S_{gf}\left(q - \frac{r}{2}\right)[Z_2\gamma_5]_{fe}\right]~,
\label{eq:BSEkernel_uchannel}
}
\end{widetext}
where now $r$ is an additional integration momentum in the BSE (cf. Eqs.\eqref{eq:inhomBSE_vector} or \eqref{eq:homogeneousBSE}). The resulting truncation of the BSE kernel and the quark DSE is shown in Fig.~\ref{fig:kernels}.

The inclusion of the two kernels given in equations (\ref{eq:BSEkernel_schannel}) and (\ref{eq:BSEkernel_uchannel}) generates a highly non-trivial analytic structure of the integrand of the BSE, induced by the intermediate pions going potentially on-shell as well as by singularities in the quark propagators, see ref.\ \cite{Miramontes:2019mco} for details where also the techniques for 
finding viable contour deformations for performing the numerical integrations in a mathematically correct way are 
described.\footnote{
Further details on applying contour deformations for performing the numerical integrations in the context 
of DSEs and BSEs can be found in refs.\ 
\cite{Santowsky:2020pwd,Windisch:2013dxa,Windisch:2014lce,Eichmann:2019dts} and references therein.}

The pion Bethe-Salpeter amplitude possesses four tensor components. Hereby, only one is generically small such that for 
a precise calculation of pion properties it would be necessary to take into account three of them, namely the leading pseudoscalar
term and two subleading ones related to the axialvector structure. However, using in the above described kernels three amplitudes 
for every pion Bethe-Salpeter amplitude in these expressions is numerically by an order of magnitude more expensive
 and far beyond the scope of the present 
study. Having restricted to the leading component of the pion amplitude in the kernels \eqref{eq:BSEkernel_tchannel} and
\eqref{eq:BSEkernel_uchannel} one can further exploit that this leading amplitude may be well approximated using the 
chiral limit value of the quark dressing function $B(p^2)$ and normalize the amplitude by dividing through the 
pion decay constant $f_{\pi}$,
\begin{equation}\label{eq:pion_chiral}
\Gamma_{\pi}^i (p;P) = \tau^i \gamma_5 \frac{B(p^2)}{f_{\pi}} \, .
\end{equation}
For light quarks the difference between calculated leading order amplitude and this approximation is at the level of a few percent,
see, {\it e.g.}, \cite{Fischer:2008wy} and references therein. Therefore we use this simplified form.

\section{Poles of the quark propagator}\label{sec:quark_poles}

In the truncations of the quark DSE used in this paper, the quark propagator features pairs of complex conjugate poles 
in the complex plane (see, e.g. \cite{Windisch:2016iud}). In order to facilitate the use of the quark propagators and 
easily identify the analytic structures generated by those poles in the form factor and vertex calculations, it is useful to parametrise the quark propagator simply as a sum of poles (see e.g. \cite{El-Bennich:2016qmb})
\begin{eqnarray}
S(p) &=& -i \Slash{p} \sigma_v(p^2) + \sigma_s(p^2) \, ,\nonumber \\
\sigma_v (p^2) &=& \sum_{i}^n \left[\frac{\alpha_i}{p^2 + m_i} + \frac{\alpha_i^\ast}{p^2 + m_i^\ast}\right]  \, ,\nonumber \\ 
\sigma_s (p^2) &=& \sum_{i}^n \left[\frac{\beta_i}{p^2 + m_i} + \frac{\beta_i^\ast}{p^2 + m_i^\ast}\right]~,
\end{eqnarray}
where the parameters $m_i$, $\alpha_i$, $\beta_i$ can be obtained by fitting the corresponding quark DSE solution 
along the $p^2$ real axis or, alternatively, on a  parabola in the complex plane that does not enclose the poles. 

In the numerical solution of the quark DSE in the complex plane we tested fits with one real and one pair of
complex conjugated poles, with two pairs of complex conjugated poles, and with three pairs of  complex conjugated poles.
Based on this tests we concluded that $n=2$ pairs of complex conjugated poles provided a precise fit, and as the use 
of three pairs of poles did not provide any further improvement, the reported calculations have been performed based on
fits with two pairs of poles.

\end{document}